\documentclass[twocolumn,aps,prl,amsmath,amssymb,showpacs,superscriptaddress,groupedaddress]{revtex4}
\bibpunct{}{}{,}{s}{}{,}
\usepackage{graphicx}   
\usepackage{psfrag}
\usepackage{dcolumn}
\usepackage{amssymb}
\usepackage{bm}
\usepackage{ifpdf}

\begin{document}

\title{Renormalization Group Flow of the Holst Action}

\author{J.-E. Daum}
\email{daum@thep.physik.uni-mainz.de}
\author{M. Reuter}
\affiliation{University of Mainz, D-55099 Mainz, Germany}


\begin{abstract}
The renormalization group (RG) properties of quantum gravity are explored, using the vielbein and the spin connection as the fundamental field variables. The scale dependent effective action is required to be invariant both under spacetime diffeomorphisms and local frame rotations. The nonperturbative RG equation is solved explicitly on the truncated theory space defined by a three parameter family of Holst-type actions which involve a running Immirzi parameter. We find evidence for the existence of an asymptotically safe fundamental theory, probably inequivalent to metric quantum gravity constructed in the same way. 
\end{abstract}

\maketitle

\section{Introduction}\label{s1}
During the past decade the gravitational effective average action \cite{mr} has been used in a number of studies trying to understand the renormalization behavior of Quantum Einstein Gravity (QEG) at a nonperturbative level. An important motivation was Weinberg's idea of Asymptotic Safety \cite{wein} according to which gravity might be nonperturbatively renormalizable and predictive if the corresponding RG flow possesses a non-Gaussian fixed point (NGFP) with a finite dimensional ultraviolet critical manifold. A quantum field theory of gravity can then be defined by performing the continuum limit there. 

For the case where the fundamental field is assumed to be the spacetime metric $g_{\mu\nu}$ the viability of this approach has been tested to some extent. All investigations carried out so far point in the direction that the RG flow of the effective average action does indeed possess a NGFP with the desired properties \cite{QEG,prop}. However, it is clear that other choices are equally plausible. In Einstein-Cartan gravity, for example, the field variables are the vielbein $e^a_{~\mu}$ and the spin connection $\omega^{ab}_{~~\mu}$, the latter assuming values in the Lie algebra of the Lorentz group. This entails an enlargement of the group of gauge transformations from ${\sf Diff}({\cal M})$ to ${\sf Diff}({\cal M}) \ltimes {\sf O}(4)_{\rm loc}$, so that gauge invariant functionals $\Gamma [e, \omega]$ constitute a new ``universality class'' different from that of metric gravity defined in terms of diffeomorphism invariant functionals $\Gamma [g]$. By augmenting the number of field components from $10$ in the case of $g_{\mu\nu}$ to $40$ for a pair $(e^a_{~\mu},\,\omega^{ab}_{~~\mu})$, Einstein-Cartan gravity in fact generalizes metric gravity; in particular, $\omega^{ab}_{~~\mu}$ can carry spacetime torsion.

Classically, the dynamics of pure Einstein-Cartan gravity is encoded in the Hilbert-Palatini action $S_{\rm HP} [e, \omega]$ which is of first order in the spacetime derivatives. Since the resulting equations of motion give rise to vanishing torsion, this functional can be regarded as the counterpart of the Einstein-Hilbert action of metric gravity. However, generic configurations $(e,\,\omega)$ contributing to the path integral underlying the effective action $\Gamma$ have non-zero torsion even if torsion should happen to vanish at the classical level: as is well-known, classical equivalence does by no means necessarily imply quantum equivalence. Therefore, the additional fields of Einstein-Cartan gravity are generally expected to crucially affect the renormalization. Moreover, the dynamics of fermions is altered by torsion already at the classical level.

The Hilbert-Palatini action can be generalized to the so-called Holst action $S_{\rm Ho} [e, \omega]$ which contains an additional term that only exists in four dimensions; its prefactor is the dimensionless Immirzi parameter $\gamma$. Since this monomial vanishes for vanishing torsion, it is absent in the metric approach. While the classical field equations are independent of $\gamma$, the corresponding quantum theory is expected to depend on it. In this respect, $\gamma$ can be compared to the $\theta$-parameter of QCD: even though the latter multiplies a topological term and therefore does not affect the classical equations of motion, observables of the quantum theory such as the electric dipole moment of the neutron are well known to depend on it. 

The Holst action lies at the heart of several modern approaches to the quantization of gravity. This includes canonical quantum gravity with Ashtekar's variables \cite{A,R}, loop quantum gravity (LQG) \cite{T}, spin foam models \cite{SF}, and group field theory \cite{GFT}. Within LQG, for instance, $\gamma$ enters the spectrum of area and volume operators as well as the entropy formula for black holes \cite{R} which exemplifies the quantum significance of $\gamma$ stated above. Furthermore, when coupled to fermions in a non-minimal way, the presence of the Immirzi term induces a {\sf CP} violating four-fermion interaction that might be interesting for phenomenological reasons, e.\,g. in the early universe \cite{laurent}. 

In LQG, $\gamma$ constitutes a fixed parameter that labels physically distinct quantum theories. However, a consistent application of the RG in the context of Einstein-Cartan gravity should treat $\gamma$ as an additional running coupling, associated with a corresponding monomial in the action functional, that is subject to renormalization in the very same fashion as Newton's constant $G$ and the cosmological constant $\Lambda$. 

Accordingly, we looked for fixed points of a Wilsonian RG flow on the theory space ${\cal T}$ made up by all functionals of $e^a_{~\mu}$ and $\omega^{ab}_{~~\mu}$ (and the required ghost fields) that respect a background-type realization of gauge invariance. Following the examples of Yang-Mills theory and metric gravity we projected this flow onto the subspace comprising functionals of the form of the bare action, i.\,e. of the Holst type. In those examples such truncations allowed for reliable investigations of the flow's UV behavior.

Our results strongly suggest the existence of two NGFPs that are {\it prima facie} both suitable for defining a quantum field theory of Einstein-Cartan gravity. They exist independently of the chosen gauge and the regularization scheme employed and therefore presumably reflect a universal feature of the full, i.\,e. untruncated flow. 

Since the underlying universality class is different from that of metric gravity, there is no reason to expect that the corresponding quantum field theories are equivalent to QEG, their metric counterpart. In particular, analyzing the fixed points of Einstein-Cartan and metric gravity in terms of truncations that happen to be equivalent at the classical level does by no means need to yield the same results. On the contrary, already the mere existence of fixed points in the $(e,\,\omega)$ universality class constitutes a novel result, completely independent of the analogue findings obtained in the context of metric gravity. 

Moreover, we find convincing evidence for a non-trivial renormalization of $\gamma$. It is not immediately obvious how to relate this result to the present understanding of $\gamma$ within the canonical approaches to quantum gravity based on the Einstein-Cartan theory.

In this Letter, we report on a first analysis of the corresponding Wilsonian RG flow of a novel type of gravitational effective average action. We briefly describe the flow equation used before presenting the results obtained. Further details will be reported elsewhere \cite{eomega2,eo-proc}.

%
%
%
\section{The Flow Equation}\label{s2}
We start out from an a priori formal functional integral ${\cal Z} = \int {\cal D}\hat{e}^a_{~\mu}\:{\cal D}\hat{\omega}^{ab}_{~~\mu}\:{\rm exp}\big\{-S[\hat{e}, \hat{\omega}]\big\}$, where the quantum fields $\hat{e}^a_{~\mu}$ and $\hat{\omega}^{ab}_{~~\mu}$ are defined on a fixed (differentiable) manifold without boundary, ${\cal M}$, and the bare action $S$ is invariant both under diffeomorphisms ${\sf Diff} ({\cal M})$ and local Lorentz rotations. We consider the euclidean form of the theory, so that the relevant group of gauge transformations is the semidirect product ${\bf G} = {\sf Diff}({\cal M}) \ltimes {\sf O}(4)_{\rm loc}$. 

For every co-frame $\hat{e}^a_{~\mu}$ and ${\sf o}(4)$-valued connection $\hat{\omega}^{ab}_{~~\mu}$ on ${\cal M}$ we are provided with an ${\sf O}(4)$-covariant derivative $\hat{\nabla}_\mu \equiv \partial_\mu + \frac{1}{2} \hat{\omega}^{ab}_{~~\mu} M_{ab}$ where $M_{ab}$ are the generators in the corresponding representation, and with the curvature and torsion tensors $\hat{F}^{ab}_{~~\mu\nu} \equiv \partial_\mu \hat{\omega}^{ab}_{~~\nu} + \hat{\omega}^a_{~c\mu}\hat{\omega}^{cb}_{~~\nu} - (\mu \leftrightarrow \nu)$ and $\hat{T}^a_{\mu\nu} \equiv \partial_\mu \hat{e}^a_{~\nu} + \hat{\omega}^a_{~c\mu}\hat{e}^c_{~\nu} - (\mu \leftrightarrow \nu)$, respectively. 

Under ${\sf O}(4)_{\rm loc}$ we have $\delta_{\rm L} (\lambda) \hat{e}^a_{~\mu} = \lambda^a_{~b} (x) \hat{e}^b_{~\mu}$, $\delta_{\rm L} (\lambda) \hat{\omega}^{ab}_{~~\mu} = - \hat{\nabla}_\mu \lambda^{ab} (x)$, while under diffeomorphisms $\delta_{\rm D} (v) \hat{e}^a_{~\mu} = {\cal L}_v \hat{e}^a_{~\mu}$, $\delta_{\rm D} (v) \hat{\omega}^{ab}_{~~\mu} = {\cal L}_v \hat{\omega}^{ab}_{~~\mu}$, where ${\cal L}_v$ denotes the Lie derivative along the vector field $v$. 

In order to arrive at a functional integral which can be computed (actually {\em defined}) by means of a functional RG flow we introduce arbitrary background fields\footnote{The background vielbein $\bar{e}^a_{~\mu}$ is assumed to be nondegenerate. It gives rise to a welldefined inverse $(\bar{e}_a^{~\mu}) \equiv (\bar{e}^a_{~\mu})^{-1}$, a nondegenerate metric $\bar{g}_{\mu\nu} \equiv \bar{e}^a_{~\mu} \bar{e}^b_{~\nu} \delta_{ab}$, and a completely covariant derivative $\bar{D} \equiv \partial + \bar{\omega} + \bar{\Gamma} \equiv \bar{\nabla} + \bar{\Gamma}$ where $\bar{\Gamma} \equiv \bar{\Gamma} (\bar{e}, \bar{\omega})$ is fixed by $\bar{D}_\mu \bar{e}^a_{~\nu} = 0$. Coordinate (frame) indices are denoted by Greek (Latin) letters.} $\bar{e}^a_{~\mu}$ and $\bar{\omega}^{ab}_{~~\mu}$, decompose the variables of integration as $\hat{e}^a_{~\mu} \equiv \bar{e}^a_{~\mu} + \varepsilon^a_{~\mu}$, $\hat{\omega}^{ab}_{~~\mu} \equiv \bar{\omega}^{ab}_{~~\mu} + \tau^{ab}_{~~\mu}$, and perform a background covariant gauge fixing leading to the functional integral 
\begin{eqnarray}
{\cal Z} &=& \int{\cal D}\varepsilon^a_{~\mu}\:{\cal D}\tau^{ab}_{~~\mu}\:{\rm exp}\big\{- S[\bar{e} + \varepsilon, \bar{\omega} + \tau] - S_{\rm gf} [\varepsilon, \tau; \bar{e}, \bar{\omega}]\big\}\nonumber\\
&& \times \int{\cal D}{\cal C}^\mu\:{\cal D}\bar{{\cal C}}_\mu\:{\cal D}\Sigma^{ab}\:{\cal D}\bar{\Sigma}_{ab} \:{\rm exp}\big\{- S_{\rm gh}\big\} \label{z-functional}
\end{eqnarray}
Here $S_{\rm gf}$ and $S_{\rm gh}$ denote the gauge fixing and corresponding ghost action, respectively, ${\cal C}^\mu$ and $\bar{{\cal C}}_\mu$ are the diffeomorphism ghosts, and similarly $\Sigma^{ab}$ and $\bar{\Sigma}_{ab}$ are those related to the local ${\sf O}(4)$. The gauge fixing is of the form
\begin{eqnarray}
S_{\rm gf} &=& \frac{1}{2 \alpha_{\rm D}\cdot 16 \pi G}\int {\rm d}^4 x\:\bar{e}\:\bar{g}^{\mu\nu}\:{\cal F}_\mu {\cal F}_\nu\nonumber\\
&& + \frac{1}{2 \alpha_{\rm L}} \int {\rm d}^4 x\:\bar{e}\:{\cal G}^{ab} {\cal G}_{ab}\label{gf}
\end{eqnarray}
where ${\cal F}_\mu$ and ${\cal G}^{ab}$ break the ${\sf Diff} ({\cal M})$ and ${\sf O}(4)_{\rm loc}$ gauge invariance, respectively. In order to ultimately arrive at a ${\sf Diff}({\cal M}) \ltimes {\sf O}(4)_{\rm loc}$ invariant effective average action we employ gauge conditions of the ``background type'' so that $S_{\rm gf} [\varepsilon, \tau; \bar{e}, \bar{\omega}]$ is invariant under the combined background gauge transformations acting on both $(\varepsilon,\,\tau)$ and $(\bar{e},\,\bar{\omega})$ while, of course, it is not invariant under the ``true'' (or ``quantum'') gauge transformations.      

Concretely, we choose both gauge conditions to be linear in $\varepsilon^a_{~\mu}$ and independent of $\tau^{ab}_{~~\mu}$ \cite{roberto-gf}: 
\begin{eqnarray}
{\cal F}_\mu &=& \bar{e}_a^{~\nu} \big[\bar{D}_\nu \varepsilon^a_{~\mu} + \beta_{\rm D} \bar{D}_\mu \varepsilon^a_{~\nu}\big]\:,\nonumber\\
{\cal G}^{ab} &=& \frac{1}{2}\bar{g}^{\mu\nu}\big[\varepsilon^a_{~\mu} \bar{e}^b_{~\nu} - \varepsilon^b_{~\nu} \bar{e}^a_{~\nu}\big] \equiv \varepsilon^{[ab]}\nonumber
\end{eqnarray}
Thus, in total, there are three gauge fixing parameters: $\alpha_{\rm D}$, $\alpha_{\rm L}$ and $\beta_{\rm D}$. 

The functional integral \eqref{z-functional} gives rise to the associated effective average action \cite{mr} in the standard way: one adds a mode cutoff to the bare action, $\Delta_k S \propto \int {\rm d}^4 x\:\bar{e}\:(\varepsilon, \tau)\,{\cal R}_k\,(\varepsilon, \tau)^{\rm T}$, couples $\varepsilon$ and $\tau$ to sources, Legendre transforms the resulting generating functional ${\rm ln}\,{\cal Z}$, and finally subtracts $\Delta_k S$ for the expectation value field in order to arrive at the running action 
\begin{eqnarray}
\Gamma_k [\bar{\varepsilon}, \bar{\tau}, \xi, \bar{\xi}, \Upsilon, \bar{\Upsilon}; \bar{e}, \bar{\omega}] \equiv \Gamma_k [e, \omega, \bar{e}, \bar{\omega}, \xi, \bar{\xi}, \Upsilon, \bar{\Upsilon}]\nonumber
\end{eqnarray} 
Therein $\bar{\varepsilon}^a_{~\mu}$, $\bar{\tau}^{ab}_{~~\mu}$ as well as $e^a_{~\mu} \equiv \langle \hat{e}^a_{~\mu} \rangle = \bar{e}^a_{~\mu} + \bar{\varepsilon}^a_{~\mu}$ and $\omega^{ab}_{~~\mu} \equiv \langle \hat{\omega}^{ab}_{~~\mu} \rangle = \bar{\omega}^{ab}_{~~\mu} + \bar{\tau}^{ab}_{~~\mu}$ denote the expectation value fields. Furthermore, we write for the ghosts $\xi^\mu \equiv \langle {\cal C}^\mu \rangle$, $\bar{\xi}_\mu \equiv \langle \bar{{\cal C}}_\mu \rangle$, $\Upsilon^{ab} \equiv \langle \Sigma^{ab} \rangle$, $\bar{\Upsilon}_{ab} \equiv \langle \bar{\Sigma}_{ab} \rangle$, and $\Gamma_k$ may be considered a functional of either the fluctuations $\bar{\varepsilon}^a_{~\mu}$ and $\bar{\tau}^{ab}_{~~\mu}$ or the complete classical fields $e^a_{~\mu}$ and $\omega^{ab}_{~~\mu}$. 

Obviously the action $\Gamma_k$ is defined on a very complicated theory space ${\cal T}$ consisting of functionals depending on two independent vielbein variables $(e,\,\bar{e})$, two spin connections $(\omega,\,\bar{\omega})$, as well as on the diffeomorphism and ${\sf O}(4)$ (anti-)ghosts. The functionals in ${\cal T}$ are constrained by the requirement of background gauge invariance. 

From the above functional integral based construction of $\Gamma_k$ one straightforwardly derives the FRGE it satisfies:
\begin{eqnarray}
\partial_k \Gamma_k = \frac{1}{2} {\rm STr} \big[(\Gamma_k^{(2)} + {\cal R}_k)^{-1} \partial_k {\cal R}_k\big]
\label{FRGE}
\end{eqnarray}
With the kernel ${\cal R}_k [\bar{e}, \bar{\omega}]$ specified appropriately, the equation indeed defines a flow on ${\cal T}$, i.\,e. it does not generate background gauge invariance violating terms. Contrary to the functional integral it (formally) derives from, it is well defined in the UV.

The derivation of the FRGE involves reinterpreting the second functional derivative of $\Gamma_k$ as an operator acting on a multiplet of fields with equal canonical dimensions. Therefore, in complete analogy with the familiar framework of iterated block spin transformations where one employs fields that are made dimensionless by suitable powers of the cutoff, and allows for a running (relative) field normalization, the field variables in (\ref{FRGE}) are understood to be rescaled according to $\bar{\varepsilon}^a_{~\mu} \rightarrow \bar{\mu}^{1/2}\bar{\varepsilon}^a_{~\mu}$, $\bar{\tau}^{ab}_{~~\mu} \rightarrow \bar{\mu}^{-1/2}\bar{\tau}^{ab}_{~~\mu}$. Here $\bar{\mu} \equiv \mu k$ has the dimension of a mass. In principle, the dimensionless factor $\mu$ may have a dependence on $k$ which controls the RG running of the relative field normalization, exactly as in the block spin case. However, since we are concerned here with the normalization of the \emph{fluctuations} rather than the background field, this is a subleading effect which is irrelevant for, and invisible in simple truncations of theory space\footnote{A running $\mu$ would occur in the analogue of a ``bi-metric truncation''. But even in the metric case the analysis of the refinements implied by this advanced class of truncations started only recently \cite{bimetric}.} such as those considered in this paper. In fact, the results to which we turn next refer to a constant $\mu$. It is reassuring to see that they are essentially independent of its precise value as long as $\mu$ is not too small; $\mu \gtrsim 2$ guarantees numerical stability.      
%
%
%
\section{Results}\label{s3}
We have solved the flow equation for $\Gamma_k$ on a three-dimensional truncated theory space spanned by actions of the Holst type:
\begin{eqnarray}
\Gamma_k &=& - \frac{1}{16 \pi G_k} \int {\rm d}^4 x\:e\,\Big[e_a^{~\mu} e_b^{~\nu}\Big(F^{ab}_{~~\mu\nu} - \frac{1}{\gamma_k} \star F^{ab}_{~~\mu\nu}\Big)\nonumber\\
&& \hspace{2.9cm} - 2\:\Lambda_k\Big] + S_{\rm gf} + S_{\rm gh} \label{holst-truncation}
\end{eqnarray}
~~In practice we used, because of the enormous algebraic complexity of the calculations involved, a slightly simplified version of the FRGE of the propertime type. An equation of the same type has been used within the Einstein-Hilbert truncation of metric gravity \cite{prop}, and virtually the same results were found as with the exact RG equation in this truncation.

The truncation ansatz \eqref{holst-truncation} consists of the Hilbert-Palatini action known from Einstein-Cartan gravity plus the Immirzi term; in fact, $\star F^{ab}_{~~\mu\nu} \equiv \frac{1}{2} \varepsilon^{ab}_{~~cd} F^{cd}_{~~\mu\nu}$ is the dual of the curvature of $\omega$, $F \equiv F (\omega)$, with respect to the frame indices. Besides $G_k$, \eqref{holst-truncation} contains two more running parameters: the cosmological constant $\Lambda_k$ and the Immirzi parameter $\gamma_k$. The gauge fixing and ghost terms are assumed to retain their classical form for all $k$, except for the replacement $G \to G_k$. The parameters $\alpha_{\rm D}$, $\alpha_{\rm L}$ and $\beta_{\rm D}$ are treated as constant in the approximation considered. Thus the truncated theory space can be coordinatized by a triple $(g, \lambda, \gamma)$ where $g_k \equiv G_k\,k^2$ and $\lambda_k \equiv \Lambda_k / k^2$ are the dimensionless Newton's and cosmological constant, respectively.  

With $t \equiv {\rm ln}\,k$, the RG equations are of the form $\partial_t g_k = \beta_g \equiv (2 + \eta_{\rm N})g_k, \:\partial_t \lambda_k = \beta_\lambda, \:\partial_t \gamma_k = \beta_\gamma$ where the anomalous dimension of Newton's constant, $\eta_{\rm N}$, and the other beta functions are given by
\begin{eqnarray}
\eta_{\rm N} (g, \lambda, \gamma) &=& 16 \pi \, g \, f_+ (\lambda, \gamma)\nonumber\\
\beta_\gamma (g, \lambda, \gamma) &=& 16 \pi \, g \, \gamma \Big[\gamma\,f_- (\lambda, \gamma) -f_+ (\lambda, \gamma)\Big]\label{flow-exact1}\\
\beta_\lambda (g, \lambda, \gamma) &=& -2\,\lambda + 8 \pi \, g \Big[2\,\lambda \, f_+ (\lambda, \gamma) + f_3 (\lambda, \gamma)\Big]\nonumber
\end{eqnarray}
The functions $f_\pm$ and $f_3$ are extremely lengthy and complicated and cannot be written down here. Parametrically, they depend on the parameters ($\alpha_{\rm D}$, $\alpha_{\rm L}$, $\beta_{\rm D}$) and $\mu$ which we keep constant.

In order to cover the neighborhood of the submanifold $\gamma = \pm \infty$ in ${\cal T}$, we introduce a new coordinate $\hat{\gamma}$. In the overlap $|\gamma| \in ~ ]0, +\infty[$ of the $(g, \lambda, \gamma)$- and the $(g, \lambda, \hat{\gamma})$-chart, the coordinates $\gamma$ and $\hat{\gamma}$ are related by the transition function $\hat{\gamma} (\gamma) = \gamma^{-1}$ so that $\beta_{\hat{\gamma}} = - {\hat{\gamma}}^2\,\beta_\gamma (g, \lambda, {\hat{\gamma}}^{-1})$. 

We studied the system \eqref{flow-exact1} and its $\hat{\gamma}$-counterpart for various cutoff functions, gauge parameters, and $\mu$ values.

The RG flow we found displays several generic features. First of all, it has an exact reflection symmetry under $\gamma \to - \gamma$. Furthermore, for $\gamma$ not too close to $\pm 1$, the functions $f_\pm$ and $f_3$ turned out almost {\em independent of} $\gamma$. For such values of $\gamma$ it is a remarkably precise approximation to replace them by functions $\tilde{f}_\pm$ and $\tilde{f}_3$ that only depend on $\lambda$, leading to
\begin{eqnarray}
\partial_t\,g_k &=& \Big[2 + 16 \pi \, g_k \, \tilde{f}_+ (\lambda_k)\Big] g_k\nonumber\\
\partial_t\,\gamma_k &=& 16 \pi \, g_k \, \gamma_k \Big[\gamma_k\,\tilde{f}_- (\lambda_k) - \tilde{f}_+ (\lambda_k)\Big]\label{flow-hypo1}\\
\partial_t\,\lambda_k &=& - 2\,\lambda_k + 8 \pi \, g_k \Big[2\,\lambda_k \, \tilde{f}_+ (\lambda_k) + \tilde{f}_3 (\lambda_k)\Big]\nonumber
\end{eqnarray}      
and likewise for the $\hat{\gamma}$-chart. The reason for the above proviso that $\gamma$ should not be close to $\pm 1$ is as follows.

\begin{figure}
	\centering
	\includegraphics[width=.9\columnwidth]{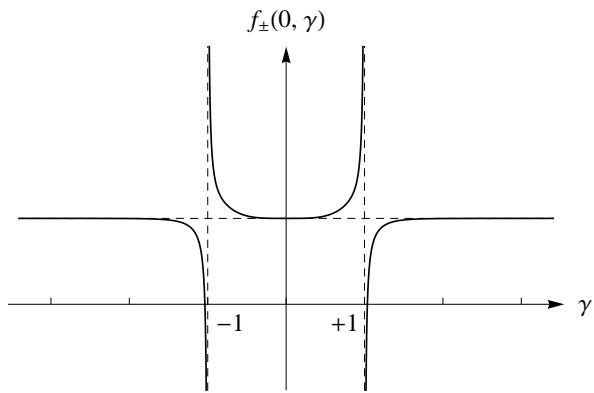}
	\caption{Schematic behavior of $f_{\pm}(\lambda=0,\gamma)$ as a function of $\gamma$. Except in a vicinity of
		$\gamma = \pm 1$, the functions are approximately constant.}
	\label{fplot}
\end{figure}

The functions $f_{\pm}(\lambda,\gamma)$ have simple poles at $\gamma = \pm 1$, but are fairly independent of $\gamma$ outside a small neighborhood of $\gamma = \pm 1$. This is a completely universal feature; it is found for all values of the gauge parameters and of $\mu$, and with all cutoff schemes employed. In Fig.\ \ref{fplot} the schematic behavior of $f_{\pm}$ in the $\lambda=0$ plane is sketched.
The singularities at $\gamma = \pm 1$ are a consequence of the fact that for these values of the Immirzi parameter the (an\-ti-) selfdual projection of $\omega^{ab}_{~~\mu}$ completely drops out from the action. Since in the functional integral equivalent to the FRGE one continues to integrate over the decoupled projection, this leads to a divergence. If one wanted to study ``chiral gravity'' based upon a selfdual connection, say, the integration over the anti-selfdual component has to be omitted, of course, and this amounts to using a new, regular FRGE, different from the one we actually analyze. Therefore, the poles at $\gamma = \pm 1$ and the zeros slightly below $\gamma=-1$ and above $\gamma=1$ are unphysical. While the equations \eqref{flow-hypo1} are certainly equivalent to \eqref{flow-exact1} when $|\gamma| \not\approx 1$, this is the reason why for $|\gamma| \to 1$, too, the regular beta functions \eqref{flow-hypo1} rather than those of \eqref{flow-exact1} are likely to apply.

The system \eqref{flow-hypo1} and its analogue in the $\hat{\gamma}$-chart imply $\beta_\gamma = 0$ and $\beta_{\hat{\gamma}} = 0$ for $\gamma^\star = 0$ and $\hat{\gamma}^\star = 0$, respectively. For each of the two sets of equations we do indeed find a fixed point $\mbox{{\bf NGFP}}_{\boldsymbol{0}} \equiv (g^\star_0, \lambda^\star_0, \gamma^\star)$ and $\mbox{{\bf NGFP}}_{\boldsymbol{\infty}} \equiv (g^\star_\infty, \lambda^\star_\infty, \hat{\gamma}^\star)$ of \eqref{flow-exact1} and the corresponding system of beta functions in the $\hat{\gamma}$-chart with $g^\star_{0, \infty} > 0$, $\lambda^\star_{0, \infty} < 0$ and $g^\star_0 \neq g^\star_\infty$, $\lambda^\star_0 \neq \lambda^\star_\infty$. This is our main result.

The discovery of these NGFPs in the new universality class based upon $e$ and $\omega$ is clearly important, a first hint at the viability of the Asymptotic Safety program in Einstein-Cartan gravity. As we pointed out in the Introduction, their existence is conceptually as well as computationally independent of, and not implied by the known properties of the metric theory, QEG.

At both fixed points, the $g$ and $\lambda$ directions are to a very good approximation eigendirections of the linearized flow on ${\cal T}$, whereas this is exactly true for the $\gamma$- and $\hat{\gamma}$-directions, respectively. At $\mbox{{\bf NGFP}}_{\boldsymbol{0}}$ and $\mbox{{\bf NGFP}}_{\boldsymbol{\infty}}$, both the $g$ and $\lambda$ directions are relevant scaling fields, they grow towards the IR and their associated critical exponents $\Theta_1$ and $\Theta_2$ are real and positive. In contrast, at $\mbox{{\bf NGFP}}_{\boldsymbol{0}}$ the Immirzi parameter $\gamma$ is irrelevant ($\Theta_\gamma < 0$), whereas at $\mbox{{\bf NGFP}}_{\boldsymbol{\infty}}$ its inverse $\hat{\gamma}$ is relevant ($\Theta_{\hat{\gamma}} > 0$).

In Table \ref{tab-NGFPprops} we display the NGFP coordinates and critical exponents for various gauge parameters $\alpha_{\rm D}$ and the fixed values $\beta_{\rm D}=0$, $\alpha_{\rm L}=16\pi g\bar{\mu}^{-4}$, and $\mu=5$. (Note that, defined as in eq. (\ref{gf}), $\alpha_{\rm L}$ is dimensionful, and that our present choice is the natural analogue of the Feynman gauge.) A comprehensive discussion of the numerical results and a careful quantitative analysis of the domain of validity of the truncation employed will be published elsewhere \cite{eomega2}.

\begin{table}
\begin{center}
\begin{tabular}{ccccccc}\hline
$\mbox{{\bf NGFP}}_{\boldsymbol{0}}$ & $g^\star_0$ & $\lambda^\star_0$ & $g^\star_0\,\lambda^\star_0$ & $\Theta_1$ &$\Theta_2$ & $\Theta_\gamma$ \\ \hline \hline
$\alpha_{\rm D} = 1$ & 3.37 & -6.78 & -22.86 & 1.94 & 3.71 & -1.98 \\
$\alpha_{\rm D} = 10$ & 1.36 & -1.08 & -1.47 & 2.46 & -6.64 & -0.43 \\
$\,\alpha_{\rm D} = 0.1$\, &\, 3.65 \,&\, -7.42 \,&\, -27.09 \,&\, 2.28 \,&\, 3.73 \,&\, -2.00 \,\\ \hline\\
\hline
$\mbox{{\bf NGFP}}_{\boldsymbol{\infty}}$ & $g^\star_\infty$ & $\lambda^\star_\infty$ & $g^\star_\infty\,\lambda^\star_\infty$ & $\Theta_1$ &$\Theta_2$ & $\Theta_{\hat{\gamma}}$ \\ \hline \hline
$\alpha_{\rm D} = 1$ & 3.30 & -4.18 & -13.79 & 1.81 & 3.22 & 1.94 \\
$\alpha_{\rm D} = 10$ & 2.18 & -1.83 & -3.98 & 2.76 & -2.40 & 1.34 \\
$\alpha_{\rm D} = 0.1$ & 3.86 & -5.16 & -19.89 & 2.55 & 3.32 & 2.01 \\ \hline
\end{tabular}
\caption{Properties of $\mbox{{\bf NGFP}}_{\boldsymbol{0}}$ and $\mbox{{\bf NGFP}}_{\boldsymbol{\infty}}$.}
\label{tab-NGFPprops}
\end{center}
\end{table} 

Defining ${\cal I} \equiv \frac{1}{16 \pi G_k} \int{\rm d}^4 x\,e\,{\varepsilon}^{\mu\nu\rho\sigma} T^a_{\mu\nu} T^b_{\rho\sigma} \delta_{ab}$, the contribution of the Immirzi term to \eqref{holst-truncation} can be written as ${\rm exp}\big\{-\frac{1}{\gamma}\cdot{\cal I} +\: \mbox{surface term}\big\}$. Therefore, for $\gamma \to 0^+$ configurations with ${\cal I} > 0$ get strongly suppressed whereas those with ${\cal I} < 0$ will be enhanced. For $\gamma \to 0^-$, the situation is just reversed. These two cases are related by parity, and neither of them leads to a {\it complete} suppression of torsion. This suggests that metric gravity is not recovered for any value of $\gamma$.

Moreover, as was already emphasized, one has to beware of taking features of specific truncations merely used in a first {\it approximate} analysis of the fixed point structure for features of the full theory; with a truncation ansatz more general than \eqref{holst-truncation} the fixed point actions will probably no longer be of the simple Holst form. Therefore, we a priori do not expect either of the two quantum field theories that presumably manifest themselves in $\mbox{{\bf NGFP}}_{\boldsymbol{0}}$ and $\mbox{{\bf NGFP}}_{\boldsymbol{\infty}}$ to be fully equivalent to QEG.

Setting  $\Lambda_k = 0$ in \eqref{holst-truncation}, we obtain the two-dimensional $(g, \gamma)$- and $(g, \hat{\gamma})$-truncation, respectively. Its analysis strongly indicates that {\it the Immirzi parameter owes its running to a non-zero cosmological constant}, i.\,e. to the presence of the associated invariant in the average action.  

With respect to variations of the regularization scheme our results are remarkably robust. The signs of the fixed point coordinates, and of similar quantities that are expected to be universal, are gauge parameter independent, as well. Nevertheless, the quantitative gauge dependence of the universal quantities such as the product $g^\star_{0, \infty} \, \lambda^\star_{0, \infty}$ and the critical exponents is stronger than in comparable calculations within metric gravity \cite{prop}.

%
%
%

\section{Conclusion}\label{s4}

We have found significant evidence for Asymptotic Safety of pure gravity in the Einstein-Cartan approach. There seem to exist two NGFPs, located at $\gamma = 0$ and $\gamma = \pm \infty$, which in principle both are suitable for taking the continuum limit there. In particular we found that the Immirzi parameter has a nontrivial RG evolution.

By investigating how observables depend upon $\gamma$, one may determine the physical properties of the resulting quantum field theories and decide which one, if any, is realized in Nature. 

Using either fixed point for the Asymptotic Safety construction, gravity is anti-screening in the UV, i.\,e. $g^\star_{0, \infty} > 0$, but in contrast to QEG the cosmological constant is negative there, $\lambda^\star_{0, \infty} < 0$ for all gauges employed. However, this does not contradict present day observations since $\lambda$ might very well flow to positive values for IR scales of the order of astronomical distances.

Future investigations should aim at a better control of the gauge dependencies and at understanding the phenomenological implications of the scale dependent Immirzi parameter. At a deeper conceptual level it will be important to understand whether and perhaps how the running $\gamma$ in the present approach can be reconciled with the constant value of $\gamma$ in LQG and similar approaches to quantum gravity. 

%
%
%
%

%
%

%

%
%
%
%
%
%
%

\end{document}